\documentclass[a4, amsmath, amssymb,pra,reprint]{revtex4-2}
\usepackage{graphicx}
\usepackage{color}
\usepackage{mathrsfs}
\usepackage{hyperref}
\usepackage{lineno}

\begin{document}

\title{Canonical Four-Wave-Mixing in Photonic Crystal Resonators: tuning, tolerances and scaling}

\author{Alexandre Chopin$^{1,2}$, Gabriel Marty$^{1,2,*1}$, In\`es Ghorbel$^{1}$, Gr\'egory Moille$^{1,*2}$, Aude Martin$^{1}$, Sylvain Combri\'{e}$^{1}$, Fabrice Raineri$^{2,3}$, Alfredo De Rossi$^{1}$}
%
\affiliation{$^{1}$ Thales Research and Technology, Campus Polytechnique, 1 avenue Augustin Fresnel, 91767 Palaiseau, France\\
	$^{2}$Centre de Nanosciences et de Nanotetchnologies, CNRS, Universit\'{e} Paris Saclay, Palaiseau, France\\
	$^{3}$Universit\'{e} Côte d'Azur, Institut de Physique de Nice, CNRS-UMR 7010, Sophia Antipolis, France\\
	$^{*1}$Now at Saint-Gobain Research Paris, 39 quai Lucien-Lefranc, 93303 Aubervilliers, France\\
	$^{*2}$Now at Joint Quantum Institute, NIST/University of Maryland, College Park, USA and Microsystems and Nanotechnology Division, National of Standards and Technology, Gaithersburg, USA\\
$^{*}${Corresponding author: alexandre.chopin@universite-paris-saclay.fr}}
%
\begin{abstract}
	Canonical Four-Wave-Mixing occurs in a resonator with only the required number of modes, thereby inhibiting competing parametric processes.
	The properties of the recently introduced photonic crystal parametric oscillator, Marty et al. Nat. Photonics, \underline{15}, 53 (2021), are discussed extensively. We compare the bichromatic design with other geometries of photonic crystal resonators. Based on a statistical study over more than 100 resonators and 10 parametric oscillators, robustness against fabrication tolerances is assessed, performances are evaluated in terms of average values and their dispersion, and the dependence on the main parameters is shown to follow the theoretical scaling. The lowest pump power at threshold is $\approx$ 40 $\mu$W and we show the existence of a minimum value of the cavity photon lifetime as a condition for parametric oscillation, which is related to three photon absorption.
\end{abstract}
\maketitle
\section{Introduction}
Non-classical states of light, e.g. entangled photons, squeezed light are ubiquitous in optical quantum sensing and quantum communication and simulation. These states are conveniently generated at room temperature through parametric down conversion in materials with second order nonlinearity\cite{kwiat1995}. Resonant enhancement in optical cavities is used to increase the efficiency of these sources, which have been miniaturized in integrated  photonic circuits. As silicon lacks second order optical nonlinearity, spontaneous Four-Wave-Mixing (FWM) is exploited as an alternative. Here, two photons from the pump decay spontaneously into a pair of photons under the constraint of energy conservation. If the interacting waves are all on resonance with the corresponding cavity modes,  the spontaneous generation rate scales as $R \propto (n_2^2 Q^3/V^2)P^2$ with $n_2$ the Kerr non-linear index, Q the quality factor, V the volume of the resonator and P the pump power \cite{helt2012,azzini2012}. Time-energy entangled photon pairs have been demonstrated on a silicon chip via FWM \cite{grassani2015} with a microring resonator. By optimizing the nonlinearity of the material and the Q factor, large efficiency can be achieved\cite{steiner2021}\\
Here we discuss a different class of resonators, photonic crystals\cite{yablonovitch1987,john1987}, which differ from ring (and racetrack, disk, ...) resonators in many ways. First, the confinement is based on Bragg scattering and not total internal reflection. Modes are spatially inhomogeneous and overlap only in part and, finally, the modal volume is at least an order of magnitude smaller than in any other dielectric resonator. Thus PhC are amenable to a very large nonlinear interaction because they enable a very strong confinement with still potentially large ($\gg1M$) Q-factors\cite{asano2017}.\\
%
\begin{figure*}[t!]
	\centering
	\includegraphics[width=0.9\textwidth]{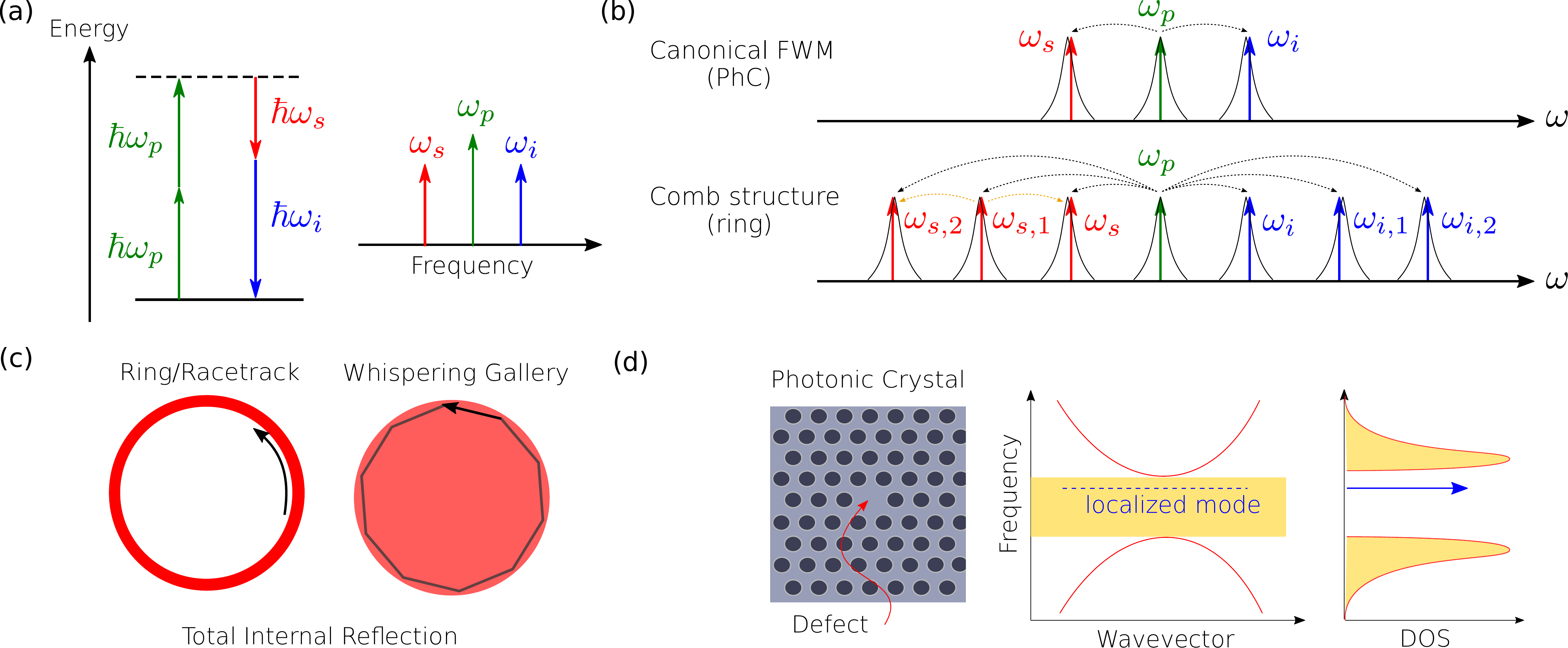}
	\caption{\label{fig:FWM} (a) : Representation of degenerate FWM: energy conservation $2\hbar\omega_p = \hbar\omega_s+\hbar\omega_i$ and idealized spectrum with the signal and the idler ($\omega_s$,$\omega_i$) symmetrically spaced relative to the pump $\omega_p$. (b) triply-resonant FWM, interacting photons (arrows) and cavity modes (black lines). Top : canonical FWM in a resonator with only three spectrally equi-spaced modes; bottom : multimode resonator with constant FSR, where multiple competing FWM processes simultaneously take place. Partly inspired from Ref. \cite{stone2022}. (c) confinement in ring, racetrack and whispering gallery resonators is due to total internal reflection; (d) a defect in a dielectric with periodic modulation (Photonic Crystal) induces a localized mode inside the forbidden bandgap (orange shaded area), a sharp peak in the Density of Optical States (DOS).}
\end{figure*}
Nanoscale devices based on Photonic crystal cavities have been demonstrated: Raman laser\cite{takahashi2013}, electrically pumped nano-laser integrated on a silicon chip\cite{crosnier2017}, pulsating Fano laser\cite{yu2017}  and all-optical memory \cite{nozaki2012}. Their common point is to operate at very low power (microWatt regime). 
The demonstration of optical parametric oscillations \cite{Marty2020} in a nanoscale PhC cavity with a threshold of $\approx$ 50 $\mu$W is particularly interesting in the context of the scalable generation of squeezed light. 
Integrated sources of squeezed light\cite{dutt2015,vaidya2020,zhang2021}, combined with a full photonic circuit\cite{arrazola2021} are used in Gaussian Boson Sampling\cite{brod2019}, a practical configuration to demonstrate quantum advantage in computing\cite{zhong2020}. While the properties of ring resonators have been extensively studied and over a variety of photonic platforms, PhC parametric sources have just been introduced and preliminary yet promising performances as sources for quantum science have been reported very recently\cite{Chopin2022}.\\
In this article we provide a detailed description of the PhC OPO physics, covering a broad range of operation and comparing a variety of devices. This work is meant to provide a comprehensive discussion of this new class of devices. In section \ref{sec:canonical_FWM} we will revisit the concept of canonical FWM, meaning FWM in a cavity allowing the interaction of only three modes (four in the non-degenerate case). We will explain why PhC are a suitable choice and how they differ from ring resonators in this respect, in particular when considering structural disorder. The complete model will also be discussed. In section \ref{sec:InGaP_PhC} we compare the properties of three geometries of PhC multimode resonators made of InGaP. In section \ref{sec:bichromatic} we report a detailed statistical analysis of a batch of new devices and show that structural disorder induces uncorrelated fluctuations of the modes of the same resonator. We also discuss the tuning mechanism in details. In section \ref{sec:OPO} we compare theory and experiment on 11 OPOs, with good agreement on threshold and slope efficiency. Here, we elucidate the reason why in InGaP PhC, parametric oscillation is possible if the Q factor is above a certain minimum.
\section{Canonical triply-resonant four-wave-mixing}
\label{sec:canonical_FWM}
Four-Wave-Mixing refers to the exchange of energy among four optical modes through the ultrafast Kerr nonlinearity. It is described as the conversion of two "pump" photons into "signal" and "idler" photons, constrained by the strict conservation of the energy; e.g. in the case of a pump-degenerate process $2\hbar\omega_p=\hbar\omega_s+\hbar\omega_i$, thus the signal and idler frequencies are located symmetrically relative to the pump, Fig. \ref{fig:FWM}a.
The resonant enhancement of FWM requires three (four in the non-degenerate case) cavity modes with frequencies with constant free spectral range, such that pump, signal and idler photons are all on resonance with the cavity. As an example, this condition is realized in a ring resonator designed to have a nearly flat dispersion. \\
Yet, for the purpose of FWM, an ideal resonator would need to have three or four modes; in contrast, ring resonators are over-moded. Let us consider a semiconductor ring resonator based on a waveguide with effective index $n_{eff}=3.0$ and group velocity $v_g=c/4.0$ in the telecom C-band spectral range ($\lambda\approx$ 1550 nm, i.e. $\nu = 193$ THz). When the free spectral range (FSR) is set to $\nu_{n+1}-\nu_n = v_g/(2\pi R) = 400 $ GHz (hence radius $R=30 \mu$m), the azimuthal order of the modes is $2\pi \nu R n_{eff}/c_0 \approx 360$. This implies that many multiple FWM interactions are allowed simultaneously, which also enables microcavity combs\cite{gaeta2019}. 
On the other hand, competition between processes is not desirable when the goal is to maximize the transfer of power from the pump to the signal and idler in an OPO (Fig. \ref{fig:FWM}b). This is discussed in recent articles \cite{stone2022,zhang2021} as well as possible strategies.
The case of an ideal triply-resonant FWM has been considered theoretically and it has been shown that, for some combination of parameters, operations are stable and the pump can be entirely converted into signal and idler \cite{ramirez2011}. 
Interestingly, it has also been pointed out that this configuration enables the manipulation of the spectral purity of a laser beam, e.g. a \textit{noise eater} \cite{matsko2019}. Let us refer to ideal triply-resonant FWM  as \textit{canonical} resonant FWM.\\
Let us therefore consider a strategy to create a resonator allowing exactly the required number of high-Q modes and with controlled frequency spacing. To this aim, let us consider a class of optical cavities which is radically different from ring, racetrack, whispering gallery resonator, as confinement is based on Bragg scattering rather than on total internal reflection (Fig. \ref{fig:FWM}c).
\subsection{Multi-mode PhC resonators}
\begin{figure}[t!]
	\centering
	\includegraphics[width=0.85\columnwidth]{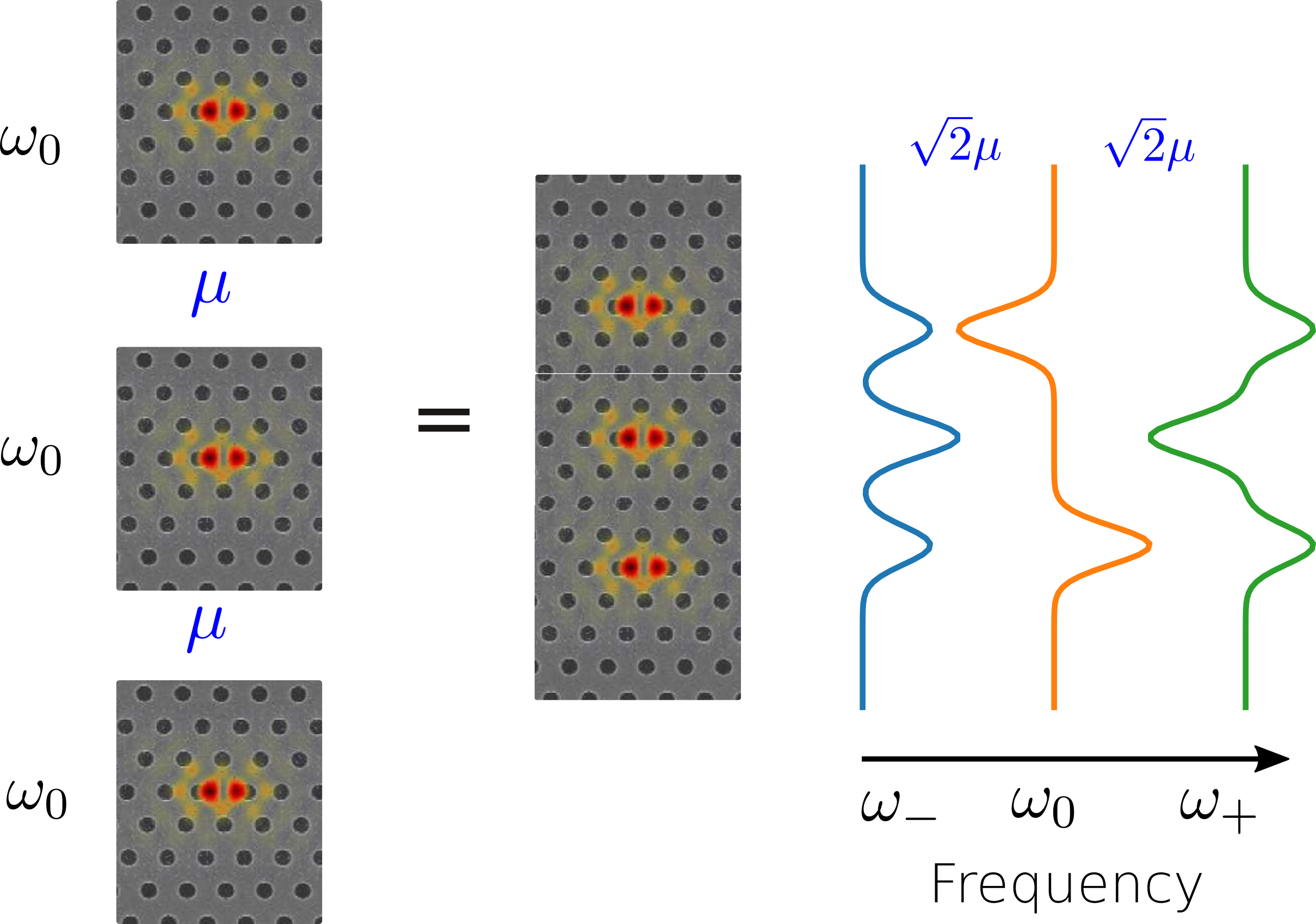}
	\caption{\label{fig:triplet}
		Triplet of modes obtained by combining three single-mode resonators with coupling strength $\mu$ and identical angular frequency $\omega_0$. They are equally split by $\sqrt{2}\mu$ in the tight-binding approximation.
	} 
\end{figure}
\begin{figure}[t!]
	\centering
	\includegraphics[width=0.9\columnwidth]{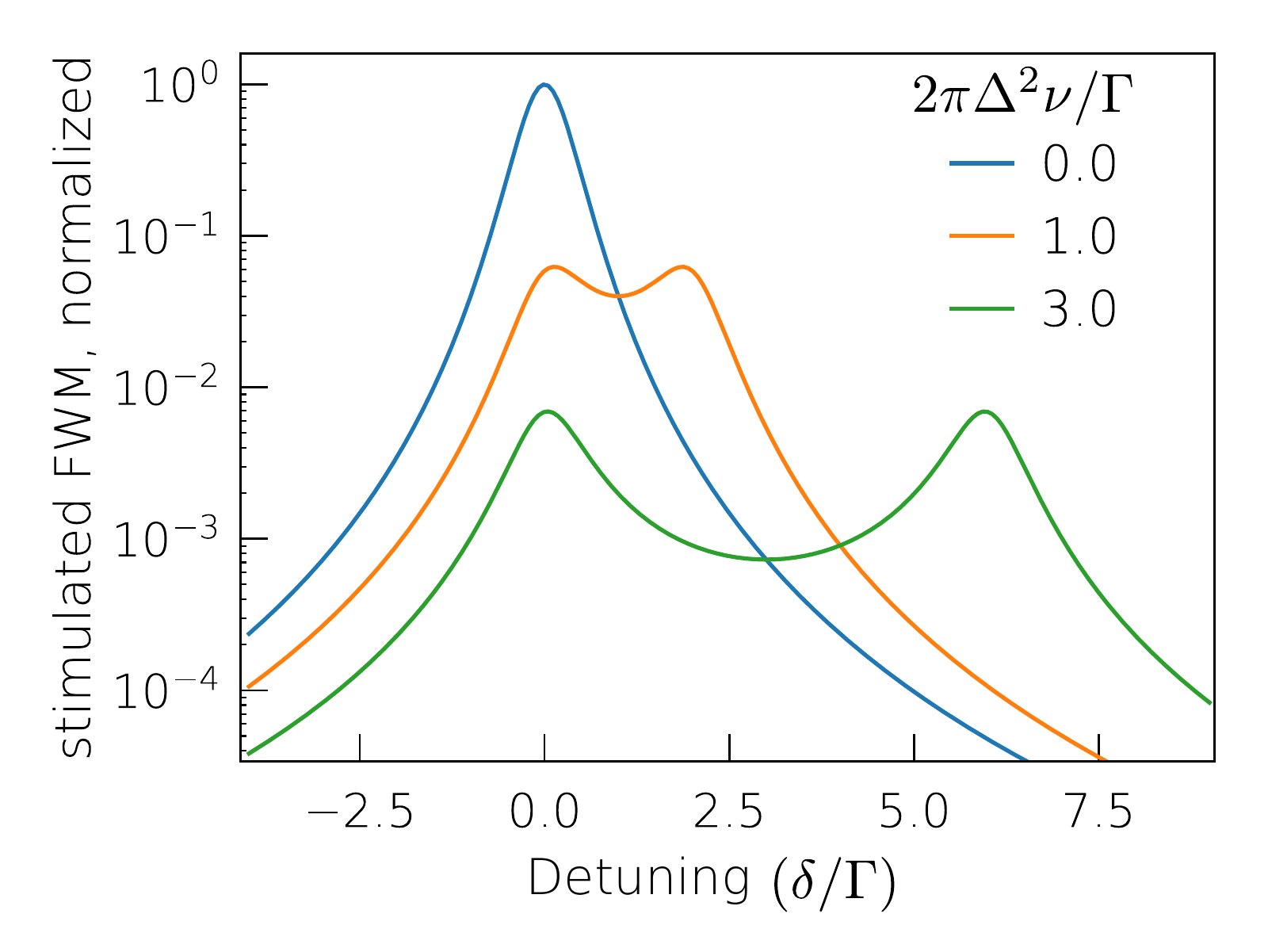}
	\caption{\label{fig:FWM_align}
		Calculated normalized stimulated FWM as a function of the normalized probe detuning, depending on the dispersion $\Delta^2\nu$.
	} 
\end{figure}

Photonic Crystals are periodic dielectric structures with complete photonic band gaps, i.e. propagation is not allowed within some spectral range in \textit{any} direction \cite{yablonovitch1987}. This condition is satisfied if the modulation of the dielectric permittivity is large enough and along any direction \cite{yablonovitch2007}. Deviations from perfect periodicity, e.g. due to disorder, result into a strong localization of light in resonant modes \cite{john1987}. Thus, modes can be created on purpose by introducing defects in the photonic crystal \cite{yablonovitch1991}, as shown in Fig. \ref{fig:FWM}d. As a consequence, a resonator with only the required modes is feasible by introducing the appropriate number of defects. Yet, there are considerable difficulties to be considered.\\
First, periodic structures with a complete band gap are very difficult to fabricate. However, it has been demonstrated that a periodic modulation in a thin slab of high index dielectric, e.g. Silicon, can be engineered to host high-Q resonances ($Q\approx 10^7$) by carefully minimizing out-of-plane radiation\cite{akahane2003,asano2017}.
The second challenge is the control of the FSR in multi-mode PhC resonators. Ultra-efficient Raman lasing was demonstrated by matching the spacing of two modes in a high-Q PhC resonator to the Raman peak in crystalline Silicon\cite{takahashi2013}. The control of more modes is increasingly challenging. Coupled PhC cavities have been considered to create multi-mode resonators aiming at dispersion control and slow-light \cite{OBrien2007,Minkov2015}, four-wave mixing and correlated photon pairs\cite{Matsuda2011,Matsuda2013},  demonstrating the optical equivalent of Electromagnetically Induced Transparency\cite{yang2009} and time-parity symmetry breaking\cite{hamel2015}.\\
According to the Tight-Binding (TB) approximation, three identical resonators coupled in a chain with strength $\mu$ create a triplet with spacing $\nu_{\pm1}-\nu_0=\pm\sqrt{2}\mu$ (Fig. \ref{fig:triplet}), which exactly corresponds to what is needed for canonical FWM. This has been experimentally demonstrated by combining three \textit{nanobeam} PhC resonators\cite{azzini2013}. Yet, the TB approximation is not adequate here: mode splitting is not symmetric because of the dispersive nature of the coupling in PhC \cite{lian2015}. Besides, this is also true in the case of ring resonators \cite{popovic2006}. Thus, even in the case of a perfect fabrication of the intended geometry, there will be a non-zero dispersion $\Delta^2\nu = (\nu_{1}-\nu_0)-(\nu_{0}-\nu_{-1})$. Triply resonant FWM is still possible if the dispersion is small compared to the spectral width of the resonance $\Gamma/2\pi$, with $\Gamma$ the cavity photon decay rate. Fig. \ref{fig:FWM_align} shows that the stimulated FWM conversion efficiency, computed using eq.\ref{eq:eta_chi_lorentz}, decreases by one order of magnitude when $|\Delta^2\nu|\approx\Gamma/2\pi$.
Thus, the consequence of misalignment is to force the choice of a lower Q factor for the resonator according to Fig. \ref{fig:FWM_align}. In Ref. \cite{azzini2013}, the Q factor is about 4000, which is low enough to ensure tolerance with respect to the dispersion and fabrication disorder. 
\subsection{Implications of the structural disorder in PhC resonators}
\begin{figure*}[t!]
	\centering
	\includegraphics[width=0.9\textwidth]{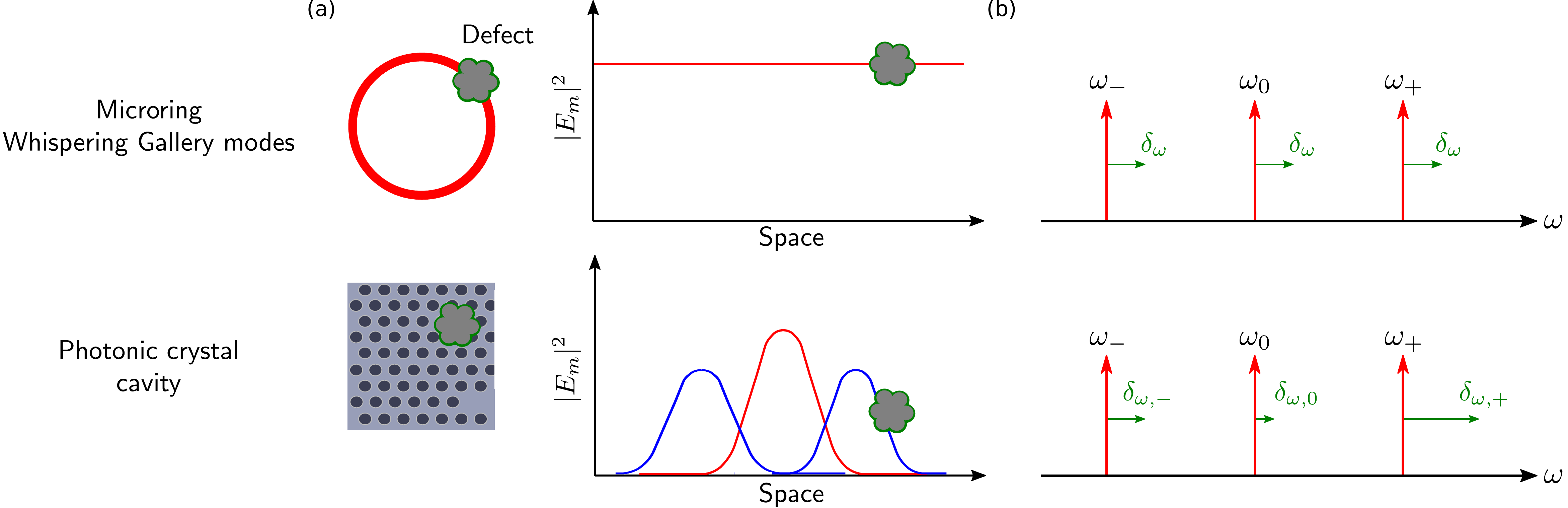}
	\caption{\label{fig:defect} Effect of a defect on the cavity eigenfrequencies. (a) : schematics of a microring resonator and a 2D photonic crystal cavity and corresponding mode field intensities in the azimuthal/longitudinal directions. The grey "cloud" represents a defect. (b) : consequence on the eigenfrequencies $\omega$ (red arrows). Horizontal green lines show the same (different) spectral shift $\delta_{\omega}$ ($\delta_{\omega,k}$) in the resonance frequencies for ring resonators (PhC) caused by the defect on the top (bottom) figure, depending on the spatial profile of the modes.}
\end{figure*}
Fabrication imperfections, e.g. surface roughness, inhomogeneities in the material, in the lithography or in the etching process, induce fluctuations in the resonances. This has been investigated in high-Q Silicon PhC resonators. Here, the standard deviation is estimated to about 40 GHz, which is considered to be a lower limit. In fact, Silicon PhC technology has demonstrated the record high Q-factor for PhC resonator and, therefore, offers state-of-the-art fabrication imperfections \cite{asano2017}. There, the standard deviation describes fluctuation of a single mode in different resonators.\\
Since modes have inhomogeneous and partially overlapping spatial distribution in a multi-mode PhC resonator, it is expected that the fluctuations of their frequencies are mutually uncorrelated. Therefore the FSR and the dispersion $\Delta^2\nu$ will essentially have the same standard deviation as the frequency of each resonance (40 GHz with state-of-the-art technology). The reason why fluctuations are uncorrelated is apparent in Fig. \ref{fig:defect}. In a ring, racetrack or microdisk resonator, the amplitude of the field of each eigenmode is uniform over the azimuthal coordinate, and it is the same for all the modes. A defect located anywhere in the cavity will induce nearly the same spectral shift over all the modes, i.e. fluctuations are correlated. The dispersive effect of defects is only visible over a large number of modes or unless it is specifically designed for, e.g. in Ref \cite{lu2014,moille2022}. \\
In contrast, in PhC multimode resonators, or coupled cavities, each mode has a well defined spatial distribution, hence, a defect will impact each mode differently, leading to weakly correlated or uncorrelated fluctuations. As a result, disorder will force the linewidth of the modes to be larger than 40 GHz, so $Q<5000$. Thus, the very high-Q achievable in PhC can only be harnessed if post-fabrication trimming or active tuning is used to compensate for the dispersion.
\subsection{Theory}
\begin{figure}[t!]
	\centering
	\includegraphics[width=0.85\columnwidth]{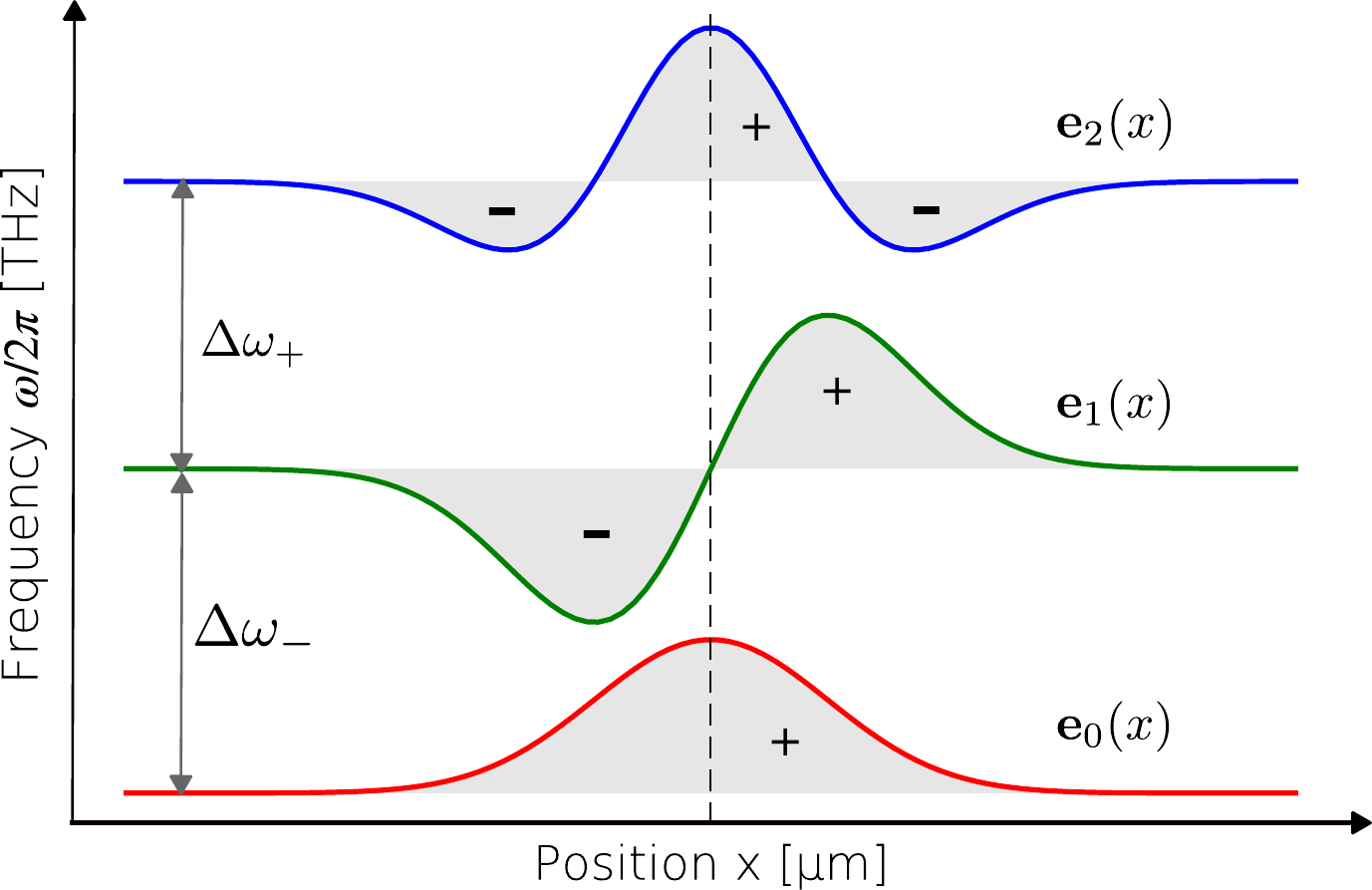}
	\caption{\label{fig:Selection_rule} Schematics of the frequency and the spatial distribution of the first normal modes $\mathbf{e_0}$, $\mathbf{e_1}$ and $\mathbf{e_2}$ in the x direction. $\Delta \omega$ denotes the FSR between successive modes.}
\end{figure}

The theory describing triply resonant FWM in our PhC cavity is based on the coupled-mode approximation\cite{manolatou1999} and it is introduced in Ref.\cite{ramirez2011}. 
Here we have slightly reformulated the model in order to highlight the role of thermo-optic effects. The electric field $\mathbf{E}$ is a superposition of non-degenerated modes $\mathbf{e}(\mathbf{r})$ with complex amplitudes $b_l$ and angular frequencies $\omega_l$: $\mathbf{E}_l(\mathbf{r},t) = b_l(t)\exp(\imath\omega_l t)\mathbf{e}_l(\mathbf{r})$. 
All the modes are assumed to have a large Q factor, which is a condition for the validity of the theory.
The master equation describes the time evolution of the complex amplitudes corresponding to the modes $l = [-,0,+]$ coupled to the waves $s_m$, i.e. propagating modes in the connecting waveguide, with angular frequencies $\omega_m$, where $m=[s,p,i]$, denoting "signal", "pump" and "idler" respectively, and subjected to the conservation of the energy: $2\omega_p = \omega_s + \omega_i$. 
The fields $s_m$ are normalized such that $|s_m|^2=P_m$ is the power carried by the wave.   
It is handy to reference the complex amplitudes $l=[-,0,+]$ to the frequencies of the corresponding inputs $m=[s,p,i]$, hence $a_l\exp(\imath\omega_m t) = b_l\exp(\imath\omega_l t)$, which defines the detuning $\delta_l = \omega_m - \omega_l$. Moreover, the fields are normalized such that $|a_l|^2=W_l$, the energy stored in the corresponding mode. Finally, the master equation reads:
\begin{equation}
	\begin{cases}
		\partial_t a_0 =& [- \imath \delta_0 -\frac{\Gamma_0}{2}]a_0 +\imath\sqrt{\kappa_0}s_0 +\\  
		&-2\imath\gamma a_0^* a_- a_+  - \frac{1}{2}\Gamma_{3PA} W_t^2 a_0\\
		\\
		\partial_t a_- =& [-\imath\delta_-  -\frac{\Gamma_-}{2}]a_- +\imath\sqrt{\kappa_-}s_s  +\\
		&  -\imath\gamma a_+^* a_0^2 - \frac{1}{2}\Gamma_{3PA} W_t^2 a_-\\
		\\
		\partial_t a_+ =& [-\imath\delta_+  -\frac{\Gamma_+}{2}]a_+  +\imath\sqrt{\kappa_+}s_i +\\
		&  -\imath\gamma a_-^* a_0^2 - \frac{1}{2}\Gamma_{3PA} W_t^2 a_+\\
	\end{cases}
	\label{eq:master}
\end{equation}
Note that the signal and idler can be exchanged and that signal-degenerate FWM is described by associating modes to inputs as follows: $0\rightarrow [s,i]$, $\pm \rightarrow p$.
The photon cavity decay rates for the modes are $\Gamma_l = \Gamma_{int,l}+\kappa_l$ with $\Gamma_{int}$ the internal decay rate (including all internal linear losses, hence absorption or scattering). The term $\Gamma_{3PA} W^2$ accounts for three-photon absorption (3PA), as two-photon absorption (TPA) is suppressed by the choice of a large gap semiconductor material. For the sake of simplicity, it is assumed that cross and self absorption rates are identical and depend on the total energy in the cavity $W_t = |a_0|^2+|a_+|^2+|a_-|^2$. The  coupling rate between the waveguide and the cavity is $\kappa$ and $\gamma$ is the non-linear coupling parameter, which is expressed in terms of $n_2$ the Kerr nonlinear coefficient and $V_{\chi}$ the effective volume describing the relevant spatial overlap between the interacting modes:
\begin{equation}
	\gamma = \frac{c_0 n_2 \omega}{\epsilon_r V_{\chi}} 
	\label{eq:gamma}
\end{equation}
with $c_0$ the speed of light and  $\epsilon_r$ the relative dielectric permittivity. The effective volume is:
\begin{equation}
	V_{\chi}^{-1}(n,m,l,k)=\frac{\varepsilon_0^2\varepsilon_r^2}{4} \int_{V_\chi}  dV
	(\mathbf{e}_i\cdot\mathbf{e}_j)(\mathbf{e}_l^*\cdot\mathbf{e}_m^*)
	\label{eq:Non_linear_volume}	
\end{equation}
with $\epsilon_0$ is the vacuum permittivity. In the case of degenerate FWM case, the subscripts of the effective volume are set as $(n,m,l,k)=(0,0,+,-)$. As apparent in Fig.\ref{fig:Selection_rule}, with this choice the integrand in eq. \ref{eq:Non_linear_volume} is non-zero if $(-,0,+)$ are adjacent modes, as it is the case in the example considered in the figure. The same holds for non-degenerate FWM, or if modes are all taken even or all odd. Thus, this is equivalent to the choice of adjacent modes in a ring resonator. We note that the effective mode volume encodes the \textit{selection rules} of the nonlinear interaction, which corresponds to the phase matching condition when FWM involves propagating waves.\\
\subsubsection{Limit of undepleted pump approximation}
Let us assume that the nonlinear interaction is weak, such that $|a_0| >> |a_-|,|a_+|$, and, therefore, the equation for the pump is unaffected by signal and idler. This decouples the system of equations \ref{eq:master} into two independent linear equations. For the pump, at steady state: $a_0 = 2\imath s_0\sqrt{\kappa_0}/(\Gamma_0+2\imath\delta_0)$, which relates the pump power to the energy in the cavity mode:
\begin{equation}
	W_0 = \frac{4\kappa_0}{\Gamma_0^2 +4\delta_{0}^2}P = \frac{4}{\Gamma_0}\mathcal{L}(\frac{2\delta_{0}}{\Gamma_0})\eta_0 P
	\label{eq:energy_power_pump}
\end{equation}
by introducing the Lorentzian function $\mathcal{L}(x) = 1/(1+x^2)$ and the escape efficiency $\eta_0$ for the pump as $\eta_0 = \kappa_0/\Gamma_0$ (and similarly for the other modes).
\\
The dynamical equations for signal and idler fields are rewritten by introducing the nonlinear coupling parameter $\chi = \gamma |a_0|^2$ :
\begin{equation}
	\left \{
	\begin{array}{rcl}
		\partial_t a_- &=& [-\imath \delta_- -\frac{\Gamma_-}{2}]a_- +\imath\sqrt{\kappa_-}s_- -\imath\chi a_+^* 
		\\
		\partial_t a_+^* &=& [\imath\delta_+ -\frac{\Gamma_+}{2}]a_+^*  -\imath\sqrt{\kappa_+}s_+^*  + \imath\chi^* a_-
	\end{array}
	\right.
	\label{eq:master_2_undepleted}
\end{equation}
\\
\subsubsection{Stimulated FWM}
The stimulated FWM is described introducing a probe field tuned to the frequency $\omega_-$ ($s_- = s_{-,in}$ and $s_+ = 0$), in eq. \ref{eq:master_2_undepleted} and solving for steady state: 
\begin{equation}
	\left \{
	\begin{array}{rcl}
		a_-/s_{-,in} &=& \sqrt{\kappa_-}(\delta_++i\Gamma_+/2)/\mathcal{G} 
		\\
		a_+^*/s_{-,in} &=& -\chi^*\sqrt{\kappa_-}/\mathcal{G} 
	\end{array}
	\right.
	\label{eq:stimu_in_out}
\end{equation}
with :
\begin{equation}
	\mathcal{G} = (-i\delta_--\frac{\Gamma_-}{2})(i\delta_+-\frac{\Gamma_+}{2})-|\chi|^2
	\label{eq:G_def}
\end{equation}
Let us consider the geometry in Fig. \ref{fig:bichromatic}a, or Fig. \ref{fig:coupl_cavity}, where the cavity is coupled to a single-ended waveguide. Then, the output and input fields are related by : $s_{\pm,out} = s_{\pm,in}+i\sqrt{\kappa_{\pm}}a_{\pm}$ \cite{combrie2017,manolatou1999}. The conversion efficiency (or probability) of the SFWM process is:
\begin{equation}
	\eta_{\chi} = \frac{|s_{+,out}|^2}{|s_{-,in}|^2} = \frac{|\chi|^2\kappa_-\kappa_+}{|\mathcal{G}|^2}
	\label{eq:eta_chi_def}
\end{equation}
From there we deduce that the maximum probability is reached at triply resonant FWM, i.e. $\delta$ = 0, with: 
\begin{equation}
	\eta_{\chi}^{max} = \frac{16 |\chi|^2\kappa_-\kappa_+}{(\Gamma_{+}\Gamma_{-} - 4|\chi|^2)^2}
	\label{eq:eta_chi_max_nl}
\end{equation}
Let us now consider eq. \ref{eq:eta_chi_def} in the limit $|\chi|\ll\Gamma$ and make explicit the dependence on the pump power by substituting for eq. \ref{eq:energy_power_pump}. This leads to:
\begin{equation}
	\eta_{\chi} = \eta_{\chi}^{max}\mathcal{L}\left(\frac{2\delta_0}{\Gamma_0}\right)^2\mathcal{L}\left(\frac{2\delta_{-}}{\Gamma_{-}}\right)\mathcal{L}\left(\frac{2\delta_{+}}{\Gamma_{+}}\right)
	\label{eq:eta_chi_lorentz}
\end{equation}
with
\begin{equation}
	\eta_{\chi}^{max} = 64\gamma^2\frac{\kappa_0^2\kappa_-\kappa_+}{\Gamma_0^4\Gamma_-^4\Gamma_+^2}P^2
	\label{eq:eta_max}
\end{equation}

\section{{InGaP} Photonic Crystal multi-mode resonators}
\label{sec:InGaP_PhC}

The PhC resonators discussed here are made of In$_{0.5}$Ga$_{0.5}$P, a III-V semiconductor alloy, lattice-matched to GaAs, with a large electronic bandgap (1.89 eV), which is more than twice the energy of photons in the Telecom spectral range. As a consequence, Two-Photon-Absorption (TPA) is suppressed, as shown in Ref. \cite{combrie2009}, also describing the fabrication process. Moreover, we have found that the residual linear absorption rate (due to defects and surface states) is extremely small, about $\Gamma_{abs} = 5\times 10^{7}$s$^{-1}$, as compared to other direct gap semiconductors \cite{ghorbel2019}. Resonators are created by altering the  position of some holes and by removing some of them. The samples have been fabricated over a time span of 6 years in TRT and C2N with comparable quality, thereby demonstrating the stability and reproducibility of the process. We will discuss three designs of multimode resonators. The first is based on a array of 10 coupled resonators, the second is a single multimode cavity based on the parabolic tapering of the lattice period and the third is a bichromatic design.

\subsection{Chain of coupled cavities}
\begin{figure}[t!]
	\centering
	\includegraphics[width=0.9\columnwidth]{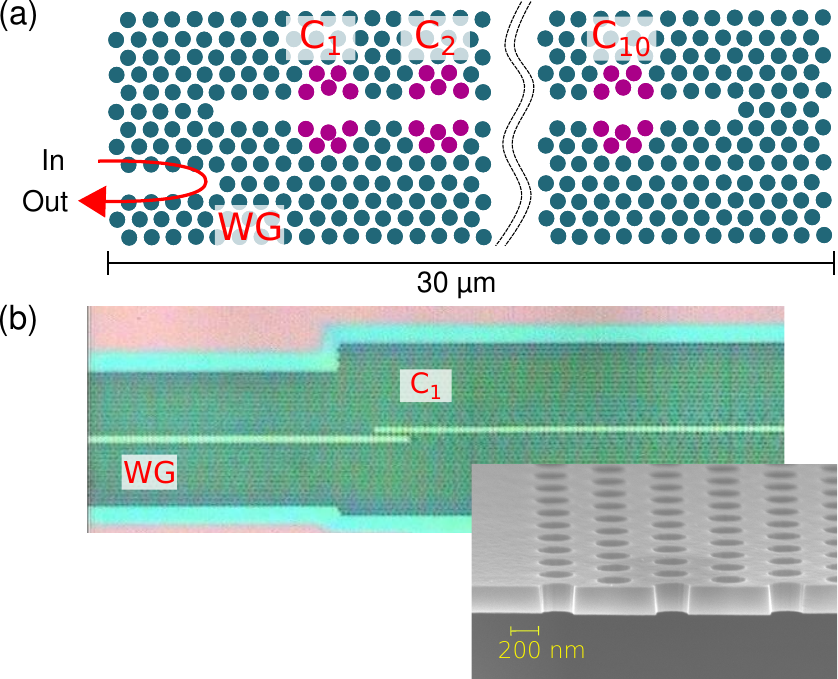}
	\caption{\label{fig:coupl_cavity} Array of coupled PhC cavities. (a), 2D layout representing the input PhC waveguide and 10 PhC cavities on a line; (b) optical and scanning electron micrograph (SEM) image revealing the suspended membrane.
}
\end{figure}
\begin{figure}[t!]
	\centering
	\includegraphics[width=0.8\columnwidth]{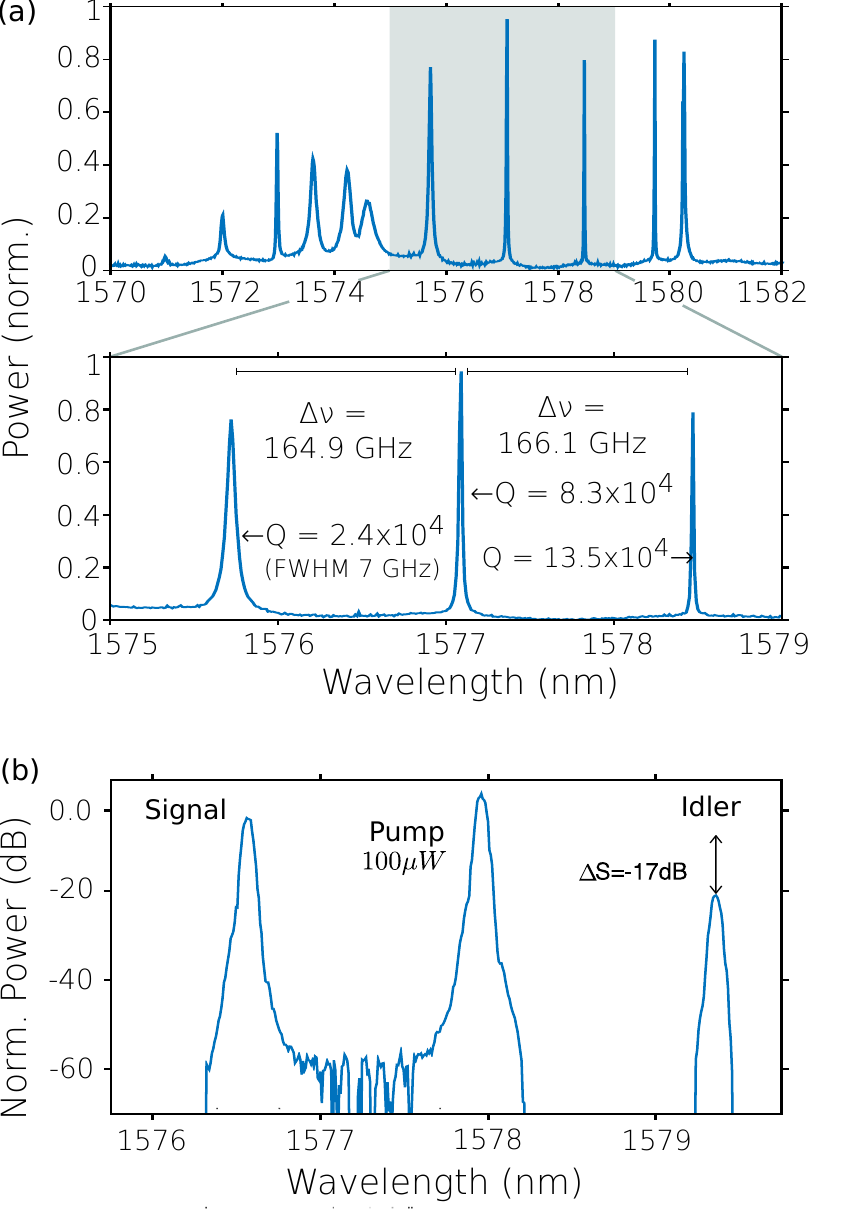}
	\caption{\label{fig:coupl_cavity_opt} (a) Linear optical scattering spectra of the chain of 10 cavities; the shadowed area is represented in detail below with frequency spacing and Q factors; (b) stimulated FWM with pump coupled to the waveguide $P_0=100\mu W$, conversion efficiency is -17 dB.
	}
\end{figure}

The chain of 10 coupled cavities is obtained by modulating the width of a photonic crystal waveguide, Fig. \ref{fig:coupl_cavity}, exactly as in ref.\cite{matsuda2012}. Each of those cavities is single mode in the spectral range of interest, therefore 10 resonances are expected due to degeneracy lift upon coupling. The linear scattering spectrum, Fig. \ref{fig:coupl_cavity_opt}a, reveals 10 peaks which, in contrast to the TB model, are not symmetrically distributed in the spectrum, in agreement with the theory including dispersive coupling \cite{lian2015}.     
Three peaks (out of 10) have been identified such that their frequency spacing is almost identical: 166.1 and 164.6 GHz; thus the difference of the FSR $\Delta^2\nu$ is smaller than the spectral width (7 GHz) of the broadest resonance, which has $Q=2.4\times 10^4$. The condition of triply resonant FWM is satisfied, as discussed earlier (Fig. \ref{fig:FWM_align}). The two other resonances have $Q=8.3\times 10^4$ and $13.5\times 10^4$.\\
The stimulated FWM in the pump degenerate configuration is demonstrated by injecting a pump with power level $P = 36$ (and 100) $\mu$W on resonance with the mode at 1578 nm, and a much weaker probe on resonance with the mode at 1576.5 nm (Fig. \ref{fig:coupl_cavity_opt}b). 
With optimized tuning of the pump and the probe, the corresponding generated idler is  --24 dB (-17 dB)  relative to the level of the reflected probe ("signal") when out of resonance; therefore the efficiency of the stimulated FWM (eq. \ref{eq:eta_chi_def}) is readily estimated to $\eta_\chi=-24 dB$ ($-17 dB$). 
Finally, assuming triply resonant FWM leads to an estimate of $\gamma=\sqrt{\eta_\chi^{max}}\Gamma_{avg}^2/16 P = (3.0\pm0.5)\times10^{22}s^{-1}J^{-1}$, from eq. \ref{eq:eta_max} in the over-coupled limit, with $\Gamma_{avg}=\Gamma_0^{1/2}\Gamma_{+}^{1/4}\Gamma_{-}^{1/4}=1.7\times10^{10}s^{-1}$. From there, the effective volume is estimated to $V_{\chi}=7\mu$m$^3$.
Raising the pump power further does not lead to an increase in the stimulated FWM efficiency. The reason will be discussed hereafter in the manuscript.
\subsection{Single nanobeam Photonic Crystal}
A semiconductor beam patterned with a single line of holes, commonly known as "nanobeam", can be engineered such as having equispaced resonances and Hermite-Gauss modes. This is achieved by tapering the period of the holes with a parabolic rule\cite{Marty2019}. Experimentally, the dispersion $\overline{\Delta^2\nu}$ is centred at zero with 20 GHz (FWHM) fluctuations, based on a statistical analysis, while the intrinsic Q factor is about $2\times10^5$ on average. Triply resonant FWM has also been demonstrated.
These resonators are fabricated on top of a silicon photonic chip through an heterogeneous integration process and are also used for nanolaser diodes\cite{crosnier2017}.
%
The properties of such resonators in the classical (stimulated FWM) and quantum regime (spontaneous FWM and correlated photons pairs) are discussed thoroughly in Ref. \cite{Chopin2022}. First, triply resonant FWM is demonstrated owing to thermal tuning.  
One advantage of the hybrid integration is that the temperature of the sample can be easily changed without affecting the coupling to the laser source. Owing to that, it is demonstrated that triply resonant FWM can be achieved by changing the sample temperature and keeping the laser wavelength fixed. Thus, much simpler laser sources can be used to drive our PhC sources. The on-chip operating power is set to about 250 $\mu$W and the maximum SFWM probability achieved is $\eta_\chi=0.18$, with averaged photon decay rate $\Gamma_{avg}=3.3\times10^{10}$ Hz ($Q\approx 37k)$. This leads to an estimate for $\gamma= 1\times10^{23}s^{-1}J^{-1}$. The effective nonlinear volume is $1.9\mu$m$^3$, which is in good agreement with the calculated value $2.2\mu$m$^3$.  
Yet, parametric oscillations have not be achieved in that sample because resonances do not have a large enough Q-factor (this will be discussed in section \ref{sec:OPO} of this paper).

\subsection{Bichromatic resonator}
\begin{figure}[t!]
	\includegraphics[width=1\columnwidth]{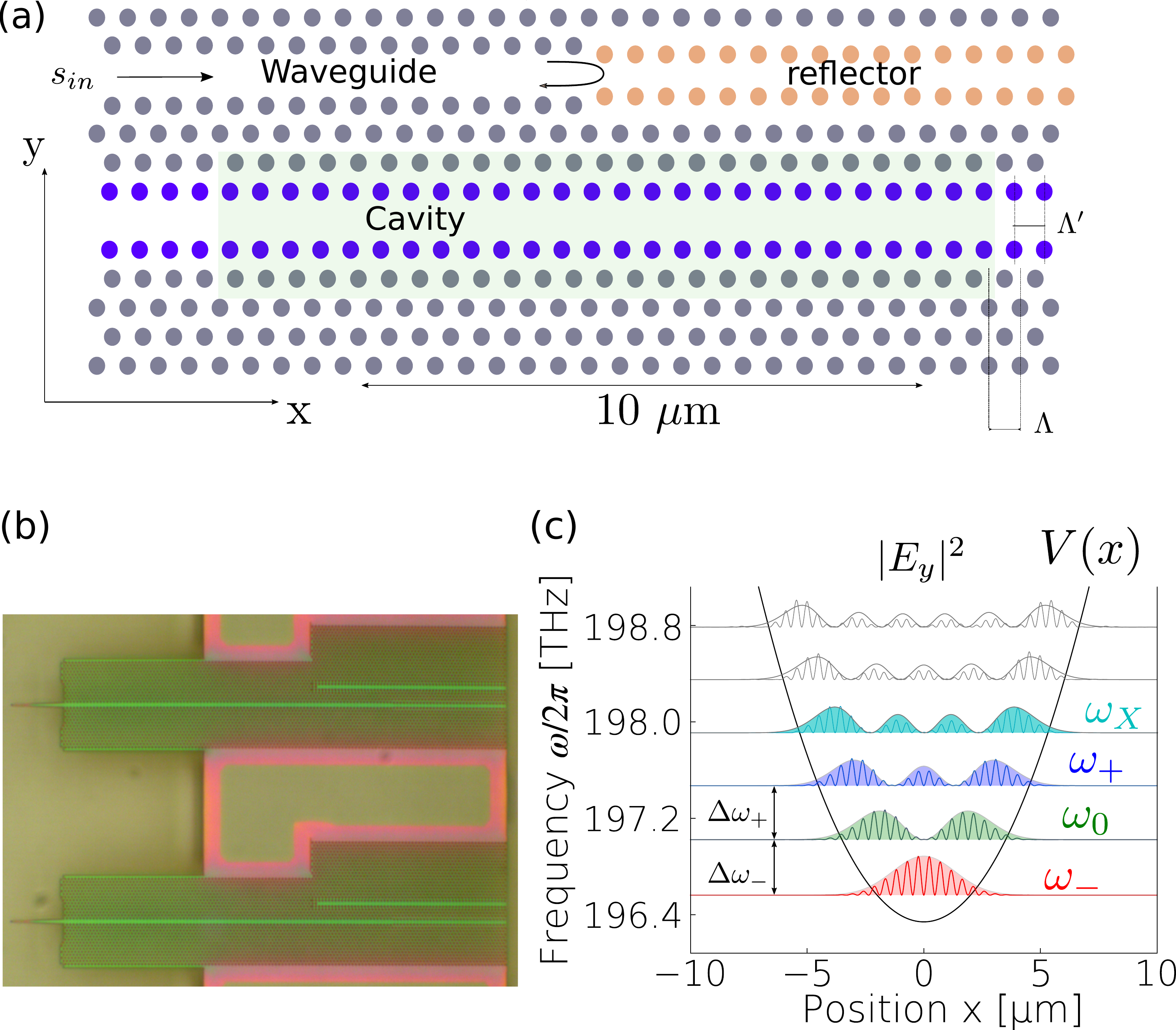}
	\caption{\label{fig:bichromatic} Bichromatic PhC cavity. (a) layout with holes with period $\Lambda$ (grey) and $\Lambda^\prime$ (blue). Input and output is provided to a single ended waveguide terminated by a reflector where holes are shifted inwards (orange); (b)  optical image; (c) calculated eigenmodes (squared field along the $x$ axis and envelopes corresponding to Gauss-Hermite functions; vertical offset corresponds to the eigenfrequencies, which are equi-spaced. The black line represents an \textit{effective} parabolic potential for the optical field. First 4 modes are denoted $\omega_-$, $\omega_0$, $\omega_+$ and $\omega_X$. See Ref. \cite{combrie2017}.}
\end{figure}
 The concept of bichromatic lattice was introduced in photonics in Ref. \cite{alpeggiani2015} and consists in a bi-dimensional lattice of holes encompassing two periods $\Lambda$ and $\Lambda^\prime$ which are reasonably close. It is shown that this leads to an effective confining potential which minimizes radiative leakage. Experimental Q factor in Silicon has been shown to be large\cite{simbula2017}. The bichromatic lattice can be implemented in different ways. In Fig. \ref{fig:bichromatic}a the period of the lattice $\Lambda$ slightly differs from the period $\Lambda^\prime=0.98\Lambda$ of the two inner lines of holes, in blue. Yet, besides nearly optimum confinement of the fundamental order mode, the bichromatic design approximates well another property, which is apparent when considering the higher order modes as well.\\ 
 This is shown in Fig. \ref{fig:bichromatic}c: the field distribution of the first 6 modes, when considering the envelope of the Bloch mode along the main axis of the cavity ($y=z=0$), correspond to the Gauss-Hermite functions and the spacing between the eigenfrequencies is constant. Thus there is a correspondence between the modes of a bichromatic photonic cavity and the states of a quantum-mechanical harmonic oscillator, whose parabolic potential is also represented in the figure and can be interpreted as an effective confining potential. This is discussed in Ref. \cite{combrie2017} where the experimental demonstration of equispaced resonances with intrinsic Q factor distributed around a most frequent value of 0.7 million is also reported.\\
The design of the cavity is completed by a single-ended waveguide formed with a missing line of holes. The input is reflected by displacing inwards the first line of holes (orange) in order to displace the cut-off frequency of the waveguide above the spectral range of interest. The position of the reflector, the displacement of the holes and the distance between the waveguide and the cavity have been optimized in order to adjust the loaded Q factor of the first modes above 0.2 million.

\section{Linear properties of bichromatic photonic crystal resonators}
\label{sec:bichromatic}
\begin{figure*}
	\centering
	\includegraphics[width=0.85\textwidth]{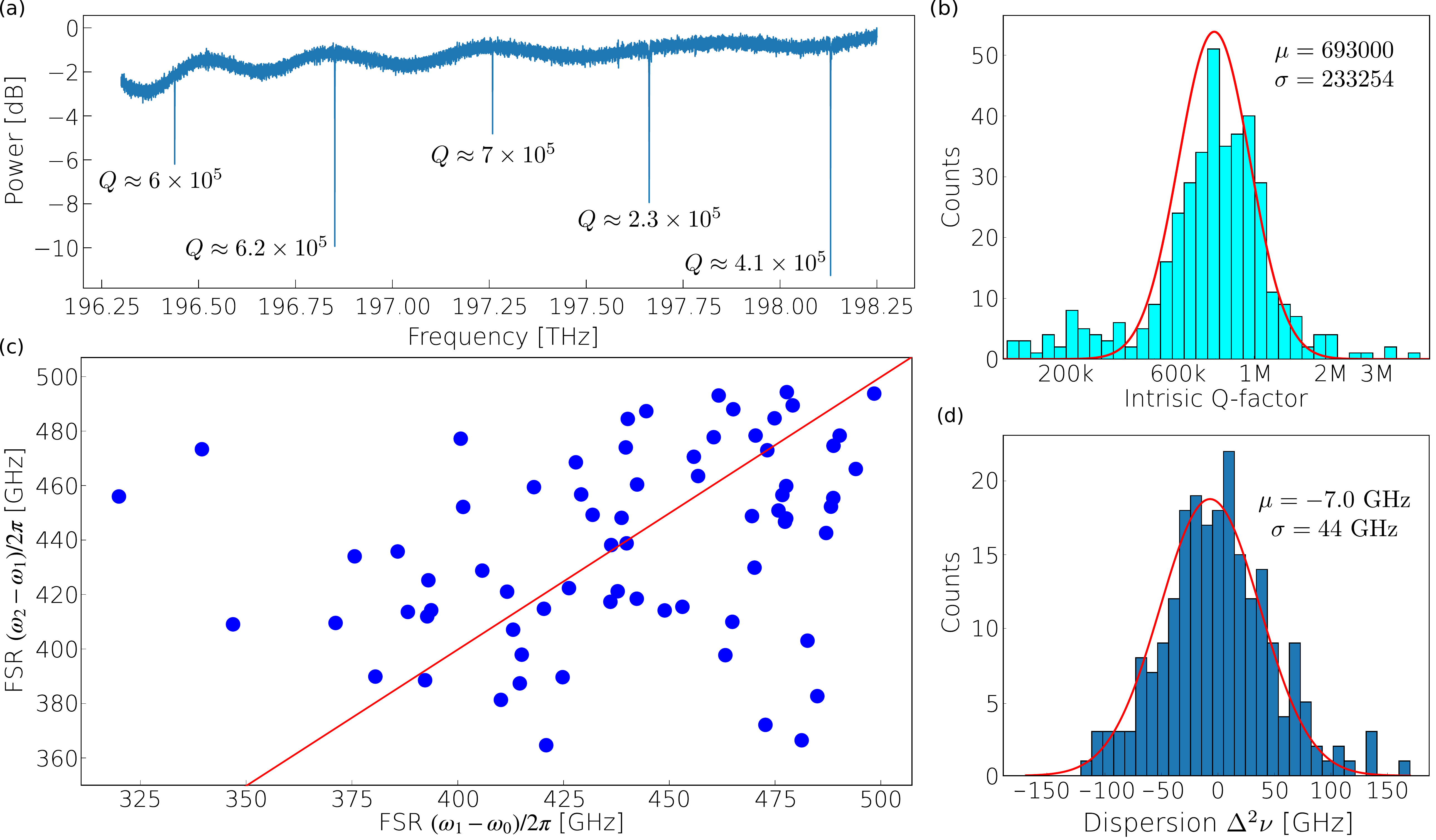}
	\caption{\label{fig:LinearCharac} Linear characterisation. (a) : Typical reflection spectrum on one of the cavity, exhibiting Q-factors with most frequent value $7 \times 10^5$. (b) : Histogram of the measured intrinsic Q-factor $Q_{int}$ of the different resonances fitted with a log-normal distribution (red line). $\mu$ is the mean value and $\sigma$ the standard deviation. (c) : FSR between mode 1 and 2 vs. FSR between mode 0 and 1, showing no correlations; red line stands for $\omega_2-\omega_1=\omega_1-\omega_0$, e.g. zero dispersion. (d) Histogram of the measured dispersion $\Delta^2\nu$ fitted with a normal distribution.}
\end{figure*}
In Ref. \cite{combrie2017} the statistical analysis of 68 bichromatic resonators concluded that: a) the intrinsic Q factor follows a log-normal distribution with most frequent value around 0.7 million and b) the average misalignment of the triplets $\Delta^2\nu$ is about 50 GHz. Several copies of the same layout have been reproduced in a different clean room with different equipments. The result is summarized in Fig. \ref{fig:LinearCharac}. 
As in Ref. \cite{combrie2017},  Optical Coherent Tomography (OCT) is used to measure the complex reflection spectrum of 124 bichromatic cavities.
From the complex reflectivity we extract the resonance frequencies (resolution 20 MHz), the intrinsic and loaded Q factors, as shown in Fig.\ref{fig:LinearCharac}a. Here, resonances are between 1490 nm  ($\approx$ 201 THz) and 1550 nm ($\approx$ 192 THz), depending on the sample. We measured loaded quality factors between $5 \times 10^4$ and up to $7 \times 10^5$, depending on the coupling between the cavity and waveguide. These new measurements confirm that the intrinsic quality factor of the resonances has a log-normal distribution peaking at 0.7 million, Fig. \ref{fig:LinearCharac}b, and reaches  values up to $1.4 \times 10^6$. Let us now consider the free spectral range (FSR). As discussed above, in a ring resonator, fluctuations of the resonances are correlated, and so as for the frequency intervals.
Let us now consider the correlations of the first $\nu_{1}-\nu_{0}$ and second $\nu_{2}-\nu_{1}$ FSR estimated for each of the resonators. Fig. \ref{fig:LinearCharac}c clearly demonstrates that fluctuations are uncorrelated, as expected. Finally, let us consider the dispersion, e.g. the difference between two successive FSR, $\Delta^2\nu$.
The corresponding histogram is fitted with a Gaussian distribution centred at  -7.0 GHz, which is much smaller than its standard deviation (44 GHz), Fig. \ref{fig:LinearCharac}d. This leads to the conclusion that the bichromatic design, on average, corresponds to equispaced resonances. 
The standard deviation is consistent with a model assuming uncorrelated fluctuations of the modes and state of the art 40 GHz fluctuation of each mode\cite{asano2017}. Thus, the probability of having triplets aligned with accuracy better than 0.4 GHz, ensuring triply resonant FWM with $Q \approx$ 0.2 million, is very low, unless post fabrication trimming is used to compensate for the fluctuations as in Ref. \cite{clementi2020}.
\\
\subsection{Active tuning}
\label{sec:Thermal_tuning}
The frequencies of the cavity modes can be tuned through the thermo-refractive effect. In the case of a multimode resonator, there are two possibilities: either all the modes will experience the same spectral shift, or, conversely, there is a differential thermo-refractive effect resulting into a change of the spacing between modes. The latter offers a tool to actively compensate for the misalignment discussed above. The field of the modes in a PhC are inhomogeneous and different, see. Fig. \ref{fig:defect}, in contrast to, e.g. ring resonators. Therefore, a thermal gradient in the cavity results into a drift of one resonance relative to another. In fact, each mode experiences a different effective temperature  $T_{m} = \int T(\mathbf{r})\epsilon(r) |\mathbf{e}_m(\mathbf{r})|^2 d\mathbf{r}$, which depends on the spatial overlap of the mode with the spatial distribution of the temperature $T(\mathbf{r})$, as discussed in Ref. \cite{sokolov2015}.\\ 
The thermal diffusion determines the spatial scale of the gradient, measured there to 5 $\mu$m (FWHM) when the InGaP PhC is surrounded by air or nitrogen at ambient pressure. 
This spatial scale is small enough to control the temperature of each cavity in the coupled chain shown in Fig. \ref{fig:coupl_cavity}, and therefore gaining reversible control on the structure of the resulting supermodes. This was demonstrated by generating a shaped distribution ultraviolet light which, upon absorption, generates a controlled profile of the temperature. The technique was based on an adaptative holographic method \cite{yu2017}. There, it is shown that the fluctuations of the resonances can be systematically compensated. Yet, if the goal is to correct small offset as in Fig. \ref{fig:LinearCharac}, the gradient induced by the dissipation of the energy in the pump mode is enough. This is a key result reported in Ref. \cite{Marty2020} and summarized here briefly. 
\subsection{"Self" thermal tuning}
\begin{figure}[t!]
	\centering
	\includegraphics[width=0.85\columnwidth]{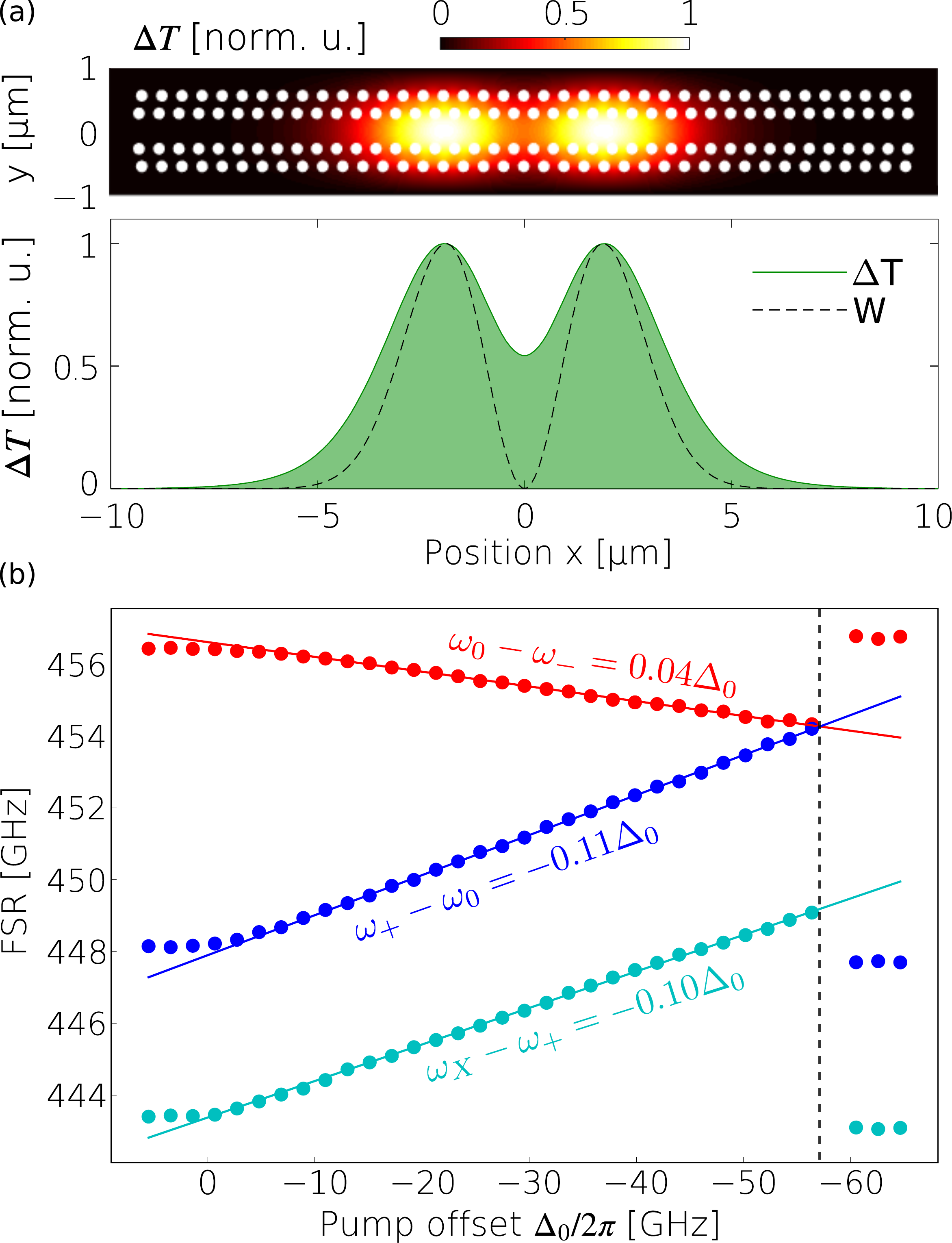}
	\caption{\label{fig:Thermaltuning} Thermal tuning (a) : Calculated distribution of the temperature change $\Delta T$ and corresponding cut along the $x$ axis (green filled curve) due to the power dissipated by mode 1 (dashed line). (b) : FSR on either side of the pump mode (blue and red circles) and linear fit (solid) vs. pump offset $\Delta_0$. $\omega_X$ is the fourth mode. Vertical dashed line represents the bistable jump.}
\end{figure}

The consequence of a thermal gradient is a drift of the modes: $\bar{\nu}_m \rightarrow \nu_m$, where the overbar refers to the sample at an uniform reference temperature (the absolute value of the temperature is not significant here). Let us consider a non-aligned triplet with dispersion $\overline{\Delta^2\nu} = \bar{\nu}_ +  + \bar{\nu}_-  -2\bar{\nu}_0$. A constant wave (CW) pump at fixed power level is tuned such that the initial positive offset $\Delta_0/2\pi = \nu_p - \bar\nu_0$ is gradually decreased. This brings the pumped mode on resonance (here, and without loss of generality, mode 1). The residual absorption rate in the material,  $\Gamma_{abs} = 5\times 10^{7}$s$^{-1}$, is enough to induce a temperature distribution modelled in Fig. \ref{fig:Thermaltuning}a. A pump-probe OCT setup is used to measure the thermal drift of all the resonances Fig. \ref{fig:Thermaltuning}b, showing that the temperature hence the dissipation linearly increases with the offset $\Delta_0$, as it changes sign. This is true if nonlinear absorption is negligible and the linewidth of the resonances is much smaller than the offset, which is the case here. Thus $W_0 \propto |\Delta_0|$.  It is apparent in  Fig. \ref{fig:Thermaltuning}b that the intervals $\nu_+ - \nu_0$ and $\nu_0 - \nu_-$ change with opposite sign as a consequence of the different effective temperature of the modes. Eventually the two intervals are equalized for some value of the offset. Interestingly, the interval with the next higher order mode will also change, but will not join the two other intervals.\\
%
The dependence of the intervals on the pump offset is accurately modeled with a linear fit $\Delta^2\nu = \overline{\Delta^2\nu} - B \Delta_0/2\pi$. As an example in figure, the fitted $B=0.15$ leads to the extrapolation of the offset $\Delta_{res}$ at which the triplet has identical spacing:   $\Delta_{res}/2\pi = \overline{\Delta^2\nu}/B = -8.3 GHz/0.15 =  -55 GHz$. We note that only negative dispersion can be compensated,  as $B>0$ due to the sign of the thermo-refractive effect. Finally, let us note that we operate in the regime of thermally induced optical bistability\cite{carmon2004}. When the cavity is exactly on resonance, which is an unstable equilibrium point, the cavity jumps to its "cold" resonance $\bar{\nu}$. This jump defines the offset $\Delta_{bist}$, which depends on the pump power available for the cavity. In the figure, the pump power is set to the value where   $\Delta_{res} = \Delta_{bist}$, let us refer to it as $P_{res}$, that is the minimal power required for triply resonant FWM. In general,  in order to achieve parametric oscillation, the pump power will have to be larger than $P_{res}$.\\
The conservation of the energy in the FWM process connects $\delta_{-}$ and $\delta_{+}$ in eq. \ref{eq:eta_chi_lorentz}. The probability of stimulated FWM depends on the pump $\delta_{0}$ and probe detuning $\delta_{-}$ and the mismatch $\Delta^2\nu$ which itself depends on the pump offset. This describes the tuning process:
\begin{equation}
	\eta_{\chi} = \eta_{\chi}^{max}\mathcal{L}\left(\frac{2\delta_0}{\Gamma_0}\right)^2\mathcal{L}\left(\frac{2\delta_{-}}{\Gamma_{-}}\right)\mathcal{L}\left(\frac{4\delta_0 - 4\pi\Delta^2\nu  -2\delta_{-}}{\Gamma_{+}}\right)
	\nonumber
	\label{eq:eta_chi_lorentz_2}
\end{equation}
Maximum SWFM probability is reached by minimizing $|\Delta^2\nu|$, so in the configuration of equi-spaced modes. Besides tuning the pump frequency, this condition can also be reached  by changing the sample temperature while the laser wavelength is kept fixed, as demonstrated recently\cite{Chopin2022}.

\section{Optical parametric oscillator}
\label{sec:OPO}
\begin{figure*}
	\centering
	\includegraphics[width=0.85\textwidth]{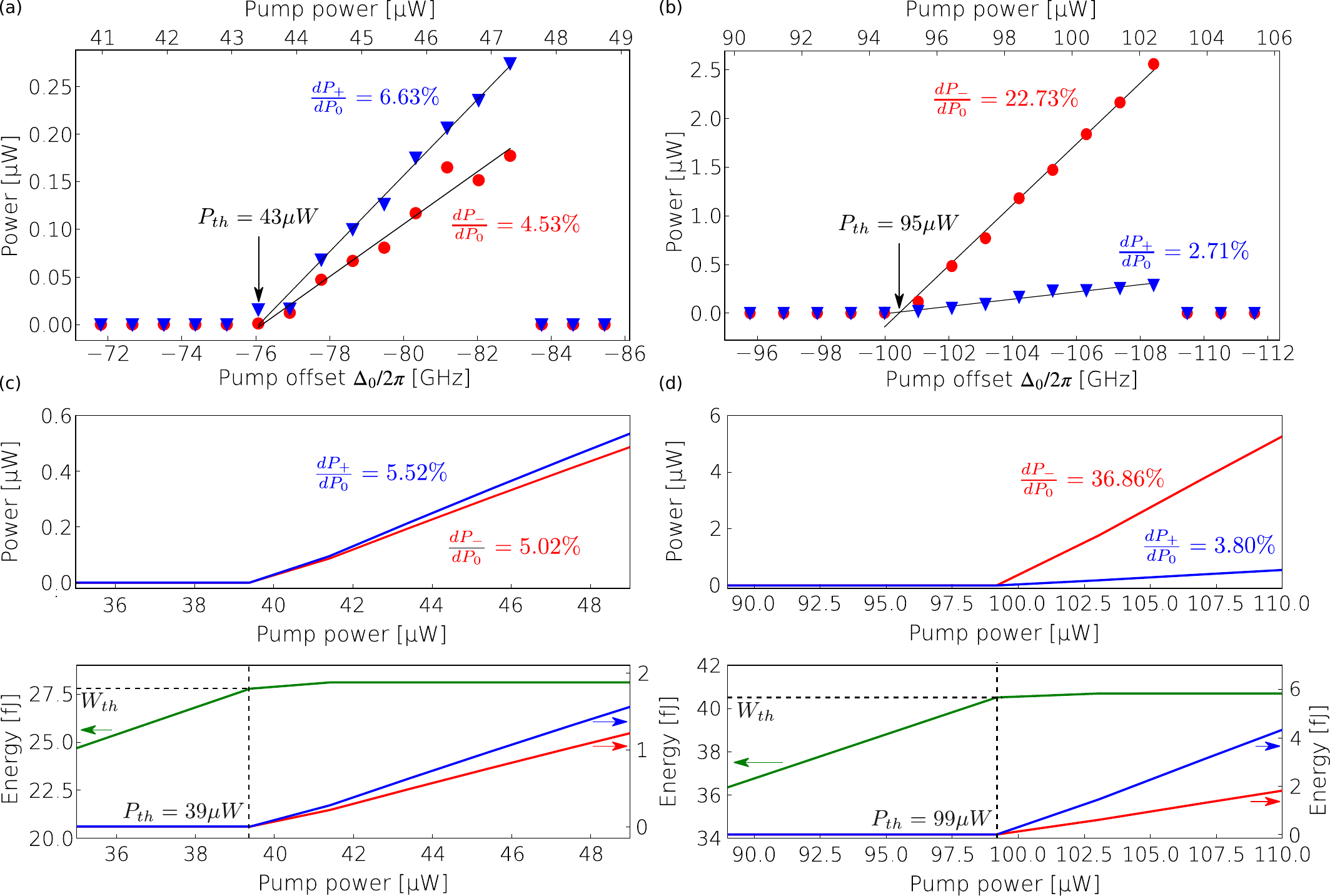}
	\caption{\label{fig:OPO_results} OPO results. (a - b)  Signal/idler on-chip power (triangles and circles) vs. the pump offset (bottom axis) and the equivalent pump power (top axis) for 2 cavities with different Q factor. $P_{th}$ is the power at threshold and $dP_{\pm}/dP_0$ are the slope efficiencies; solid lines are a fit. (c - d) Calculated output power (top) and energy (bottom) stored in the three modes: idler (blue), pump (green) and signal(red), vs. pump power for the cavities whose parameters correspond to the experiments (a) and (b). $W_{th}$ is the energy at threshold.}
\end{figure*}
We report here the study on a new batch of PhC OPO resonators, with the purpose to confirm the expected trends on quality factors of the triplet $(Q_-, Q_0, Q_+)$ and their dispersion $\overline{\Delta^2\nu}$. Thus, out of the 124 resonators considered in the previous section, about one half have the "wrong" dispersion (the statistics is almost centred in 0) and cannot be aligned, as discussed above. Yet, not all of the resonators have reached the threshold for parametric oscillation. In Ref. \cite{Marty2020} we have observed that OPO is reached only if Q factor is large enough. Here we carry a detailed study to clarify the reason.\\
We consider OPO operation on eleven resonators. The procedure is the same as in  Ref. \cite{Marty2020}. The pump laser is tuned in order to align the three modes while the output is acquired by an optical spectrum analyser. As shown in Fig.\ref{fig:OPO_results}a-b, at a given value of the pump offset, the on-chip power in the side modes abruptly increases, proving optical parametric oscillation. It is possible to define an equivalent pump power in the cavity $P_{0}$ using the pump offset, its value at which the bistable jump occurs and the power in the waveguide $P$ as $P_{0} = (\Delta_0/\Delta_{bist})P$. Then, it defines the power at threshold $P_{th}$ for parametric oscillations. The minimum measured value here is 43 $\mu$W. Above threshold, the output power evolves linearly with $P_0$ defining the slope efficiency $dP_{\pm}/dP_0$, Fig.\ref{fig:OPO_results}a-b.
\subsection{Comparison with theory}
We have modeled the experiment by solving the master equation (see Eq.\ref{eq:master}) with the measured values for $(Q_-, Q_0, Q_+)$, loaded and intrinsic, and  assuming triplet alignment and non-linearity as in Ref. \cite{Marty2020}, namely $\gamma=4\times10^{22}s^{-1}J^{-1}$ for the triplet involving the fundamental order mode and $\gamma=4.7\times10^{22}s^{-1}J^{-1}$ for the others.  Measurements on samples with low Q factor and in a regime with low efficiency, where the gain is expected to be linear, can be used to deduce a value for $\gamma$, using again eq. \ref{eq:eta_max} in the limit for overcoupled modes and triply resonant FWM: $\gamma=\sqrt{\eta_\chi^{max}}\Gamma_{avg}^2/16 P = 6 \times10^{22}s^{-1}J^{-1}$, from  in the over-coupled limit, with $\Gamma_{avg}=\Gamma_0^{1/2}\Gamma_{+}^{1/4}\Gamma_{-}^{1/4}=2.0\times10^{10}s^{-1}$ and $\eta_\chi^{max}=0.008$. The estimated $\gamma$ is close although a bit larger than calculated.\\
Let us consider two cases of OPO in Fig.\ref{fig:OPO_results}, which mainly differ in the loaded Q factor. In panel (a) we show the case of the lowest pump threshold, 43 $\mu$W and also almost balanced signal and idler. Here the loaded Q factor of signal and idler is among the largest observed. In panel (b) we show another case of higher threshold and unbalanced output. The theory above, with the parameters corresponding to the two devices,  reproduces well the results, as apparent in Fig.\ref{fig:OPO_results}c - d. More generally, the agreement is systematic, as discussed hereafter; we also note that the slope efficiency is also well reproduced, with a few exceptions. This might be due to an inexact estimate of the internal losses. Interestingly, the model shows the clamping of the energy in the pump mode when above threshold.
\subsection{Scaling with parameters}
\begin{figure}
	\includegraphics[width=1\columnwidth]{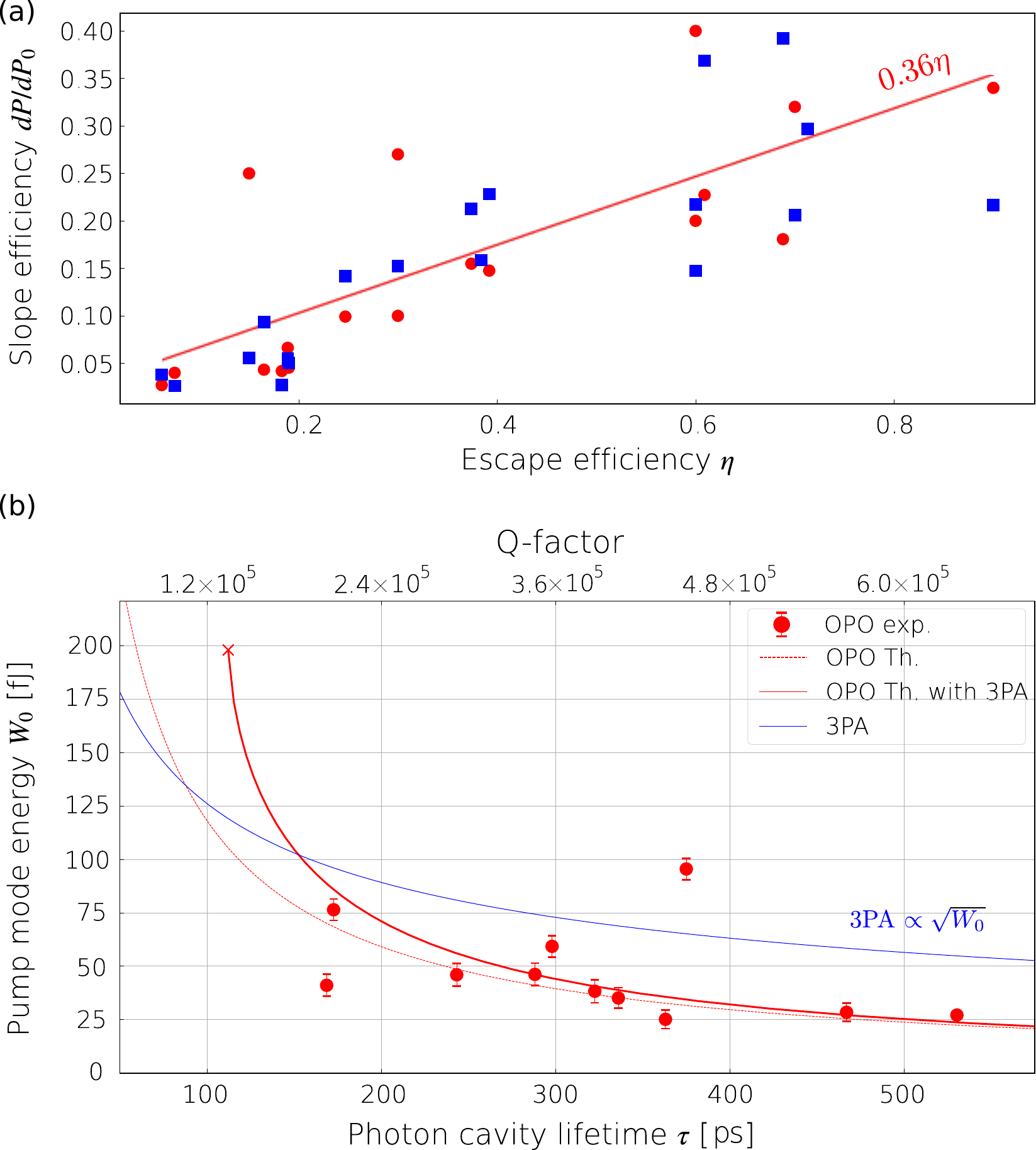}
	\caption{\label{fig:OPO_scaling} Scaling rules for OPO. (a) Slope efficiency $dP_\pm/dP_0$ vs. escape efficiency $\eta_\pm$, experiment (red circles) and theory (blue squares) and linear fit of the experimental data. The fit of the theoretical points is very close $dP/dP_0=0.35\eta$. (b) Measured energy in the pump mode $W_0$ at threshold vs. the averaged photon lifetime of the side modes $\tau^{-1} = \sqrt{\Gamma_{+}\Gamma_{-}}$, red symbols with error bars, compared with the theory, red line, with (solid) or without (dashed) including three photon absorption; the symbol "X" denotes the bifurcation of the OPO threshold (no OPO for lower $\tau$). Blue line denotes the scaling of $W_{3PA}$ with $\tau$, deduced using eq.\ref{eq:3PA_equation} and values that can be found in the Appendix.
	}
\end{figure}

The measurements carried out on 11 OPOs allow us to better assess our model and to attempt a verification of the main scaling rules. The OPO all slightly have different parameters, mainly the cavity to waveguide coupling and the detuning.
Let us first consider the slope efficiency for signal and idler, Fig.\ref{fig:OPO_scaling}a.
From the theory, neglecting nonlinear absorption, the slope efficiency is related to the escape efficiency: $P_{\pm} = \eta_{\pm}(P_0-P_{th})/2$. Thus we expect a correlation between escape and slope efficiency, which is clear in the figure, although deviations from the linear fit are not small. Moreover, the fit coefficient is 0.36 instead of 0.5. We have used the theory to generate the same plot and confirmed the dispersion of the results. Moreover, the theoretical slope, 0.35, is also very close to the measurements. This is likely to be due to nonlinear absorption, as well as to some dynamical effects, since eq. \ref{eq:master} is solved in the time domain and the input-output dependence is extracted assuming steady state.
The OPO pump energy threshold is deduced from eq. \ref{eq:G_def}, setting $|\mathcal{G}|=0$ and, when the triplet is aligned, gives: 
\begin{equation}
	W_{th} = \frac{\sqrt{\Gamma_{+}\Gamma_{-}}}{2\gamma}
	\label{eq:W_th}
\end{equation} 
Introducing the signal/idler photon cavity lifetime $\tau = (\Gamma_{+}\Gamma_{-})^{-1/2}$ leads to $W_{th} \propto \tau^{-1}$. Furthermore, using eq. \ref{eq:energy_power_pump} and eq. \ref{eq:gamma} gives:
\begin{equation}
	P_{th} = \frac{\epsilon_r V_{\chi} \omega}{8 c_0 n_2}\frac{\omega^2}{\Gamma_{avg}^2}\frac{1}{\eta_0}
	\label{eq:P_th}
\end{equation}
with $\Gamma_{avg}$ defined above.\\
Fig.\ref{fig:OPO_scaling}b shows the pump energy at threshold,  deduced from the measured $P_{th}$ using Eq.\ref{eq:energy_power_pump}, as a function of the photon cavity lifetime.  First, let us compare the experimental points with eq. \ref{eq:W_th}, which does not consider nonlinear absorption. The expected dependence is reproduced with significant deviations. There are many reasons for that: the uncertainties in the intrinsic Q or on the fact that we might have used a higher order triplet, where $\gamma$ is expected to be slightly larger than for the triplet starting from the fundamental order mode. Moreover, we do not account for the subtleties of the thermal tuning and that the power at threshold is extrapolated. In fact, a fixed power is sent to the sample, in order to tune the triplet, as discussed before. Yet, if  $P_{res}>P_{th}$, which depend on two unrelated parameters, $\overline{\Delta^2\nu}$ and $\tau$, then the minimal threshold power is not accessible experimentally.\\
\subsection{Nonlinear absorption vs. cavity photon lifetime}
The most striking fact is that we could not observe OPO, regardless the pump power, for any triplet with photon lifetime  shorter than 150 ps. In order to explain that, we have calculated the energy threshold including nonlinear absorption, from the implicit equation:
\begin{equation}
		\Gamma_{3PA} W^2 + (\Gamma_{TPA} - 2\gamma) W  + \sqrt{\Gamma_{+}\Gamma_{-}}=0
		\label{eq:3PA_equation}
\end{equation}
Under the experimental condition, two-photon absorption is vanishing, $\alpha_{TPA}=0$ and the three photon absorption coefficient is set to $\Gamma_{3PA}=2\times10^{35}s^{-1} J^{-2}$, see Appendix for details.
The solution of the implicit equation is plot in Fig.\ref{fig:OPO_scaling}b: the threshold diverges for $\tau  =(\Gamma_{+}\Gamma_{-})^{-1/2}\approx 110$ ps and no threshold exists for lower photon lifetimes. The interpretation is straightforward: the energy threshold scales as $1/\tau$ while the energy at which three photon absorption becomes dominant, namely  $\Gamma_{3PA}W^3 = \Gamma W$, is $W_{3PA} = \Gamma_{3PA}^{-1/2} \Gamma^{1/2}$, shown in Fig. \ref{fig:OPO_scaling}b, which scales as  $\tau^{-1/2}$. Therefore, OPO is possible if $W_{3PA}>W_{th}$, which is apparent in the figure.\\
We note that in silicon, the main nonlinear contribution is two-photon absorption (2PA) that scales as $\tau^{-1}$. Here the condition is simply $\Gamma_{TPA} - 2\gamma<0$, which cannot be satisfied due to its unfavourable nonlinear figure of merit. Thus OPO is not possible in Silicon.
\section{Conclusion}
Photonic crystal multimode cavities now offer a way to generate light through parametric conversion. They naturally implement \textit{canonical} FWM, which means that the resonator contains a minimal number of modes, and that three of them can be tuned into triply-resonant FWM. Tuning relies on the dissipation of the pumped mode and the ensuing thermo-refractive effect; hence we refer to it as \textit{self-tuning}.  This entails the following properties: first of all, competing parametric processes are suppressed because they are not resonant; additionally, the size of the resonator (roughly related to the number of modes) is highly reduced, which implies larger nonlinear coupling.\\
We have carried out a thorough characterization of more than 100 resonators to evaluate main parameters (internal losses, coupling, dispersion) in terms of their \textit{statistical} distribution, which gives a fair measure of the robustness and reproducibility of this technology.  We have chosen 11 cavities for analyzing the pump power threshold and slope efficiency. We confirm that they are well related to the parameters of the resonator. In particular, the slope efficiency is well correlated to the escape efficiency, that is the ratio between coupling and total losses. Moreover, the power at threshold is strongly dependent on the photon lifetime of signal and idler modes and we find a lower limit to the photon cavity lifetime, roughly corresponding to $Q = 1.3\times10^5$, below which parametric oscillation is not possible. This is because of three-photon absorption. The lowest threshold is $\approx$ 40 $\mu$W corresponding to a large Q-factor of $\approx 6 \times 10^5$.\\ 
We have compared the 2D bichromatic design with alternative approaches: coupled cavities, also based on a 2D photonic lattice, and the III-V Silicon hybrid nanobeam PhC.  All of them demonstrated resonant FWM and have comparable nonlinear coupling (in the nanobeam it is about two-fold larger than in the bichormatic, and substantially larger than for the coupled cavities) but we could not observe OPO in the two other structures, likely because the Q factor was not large enough.\\
We have considered a few designs of multimode resonators suitable for resonant FWM. We recently developed a general tool to achieve exactly equispaced modes in a generic PhC resonator\cite{Talenti2022}, under the condition of "gentle" confinement\cite{akahane2003}, implying tapered photonic structures. This tool could be used to broaden the range of geometries amenable to create OPO and other ultra-efficient parametric devices.   
This, the combination of low threshold power, high slope efficiency, and low footprint (120 $\mu m^2$ in our case) of our systems is an asset for integrated quantum optics, in particular for the generation of squeezed light via canonical degenerate FWM on chip. 

\section*{Appendix : Nonlinear absorption}
We derive the three photon absorption generation rate using the same method as in Ref. \cite{moille2016}, eq 13. The power dissipated in the cavity is expressed in terms of the 3PA coefficient $\alpha_{3PA}$ ($m^2/W^2$), which is instead defined for propagating waves, namely:
\begin{equation}
	\frac{dI}{dz} = -\alpha_{3PA} I^3 
\end{equation}
where the irradiance is related to the energy density $\mathcal{W}$ through $I = c_0 \mathcal{W}/n$ in an uniform medium of index $n=\sqrt{\varepsilon_r}$. Here below we establish the connection. 
Noting that $\frac{dI}{dz} = \frac{d\mathcal{W}}{dt}$, this can be rewritten as:
\begin{equation*}
	\frac{d\mathcal{W}}{dt} = -\alpha_{3PA} \mathcal{W}^3\frac{c_0^3}{n^3} 
\end{equation*}
this equation holds locally in any isotropic non dispersive medium. Integrating over the volume containing the cavity and using the definitions for the normal cavity modes defined before:
\begin{equation*}
\begin{split}
	\int_V \frac{d\mathcal{W}}{dt} dV=-\int_V{\alpha_{3PA}\left(\frac{c_0\epsilon_0\epsilon_r|\mathrm{e}|^2}{2n}\right)^3dV}=
	\\
	= -\alpha_{3PA}\left(\frac{c_0}{n}\right)^3|a|^6\int_V{\frac{\epsilon_0^3\epsilon_r^3}{8}|u|^6dV}
\end{split}	
\end{equation*}

This can be rewritten as: 
\begin{equation}
	\frac{d|a|^2}{dt}=-\alpha_{3PA}\left(\frac{c_0}{n}\right)^3\frac{|a|^6}{\left(V_{3PA}\right)^2}=-\Gamma_{3PA}|a|^6
\end{equation}
with the r.h.s. term $\Gamma_{3PA}|a|^4$ representing nonlinear absorption rate. This defines the three photon absorption volume: 
\begin{equation}
	V_{3PA}^{-2} = \int_V{\frac{\epsilon_0^3\epsilon_r^3}{8}|u|^6dV}
	\label{eq:V3PA} 
\end{equation}
and the nonlinear absorption coefficient:
\begin{equation}
	\Gamma_{3PA} =  \frac{\alpha_{3PA}}{\left(V_{3PA}\right)^2}\left(\frac{c_0}{n}\right)^3
	\label{eq:gamma3PA} 
\end{equation}
%

The material coefficient $\alpha_3=2.5\times10^{-26}$m$^3$W$^{-2}$ is measured in Ref. \cite{dave2015} and the nonlinear volume is calculated to $V_{3PA}=3.5\times10^{-19}$ m$^3$, which is approximately constant for all modes except for the fundamental. This formula leads to the estimate for $\Gamma_{3PA}=2\times10^{35}s^{-1} J^{-2}$ which is taken for the modeling.

\section*{Acknowkedgement}
We acknowledge financial support from the Agence Nationale de la Recherche (ANR) under the contract COLOURS (ANR-21-CE24-0024).
We are indebted with Allard Mosk, Jin Lian , Sergei Sokolov, Karindra Perrier, Ga\"elle Lehoucq for the fruitful collaboration in the context of the European Research Council (ERC), project PHAROS (grant 279248, P.I A. P. Mosk).
A. Chopin acknowledges support from IDF Quantum Saclay, 2021, A. Martin acknowledges support from IDEX Paris Saclay - IDI 2013. A. de Rossi and S. Combri\'e thank A. Matsko (OEwaves) for enlightening discussions on the parametric interactions.
\bibliographystyle{naturemag}


\end{document}